\documentclass[journal]{IEEEtran}

\usepackage{xcolor}
\usepackage{bm}
\usepackage{cite}
\usepackage{amsmath,amssymb,amsfonts}
\usepackage{algorithmic}
\usepackage{graphicx}
\usepackage{textcomp}
\usepackage{xcolor}
\usepackage{bm}
\usepackage{breqn}
\usepackage{amssymb}
\usepackage{booktabs}
\usepackage{url}
\usepackage{amsmath,graphicx}
\usepackage{color,soul}
\usepackage{bm}
\usepackage{amssymb}
\usepackage{amsmath,graphicx}
\usepackage{color,soul}
\usepackage{bm}
\usepackage{tikz}
\usepackage{amssymb}
\usepackage{latexsym}

\begin{document}

\title{Deep Adaptive Input Normalization \\for Time Series Forecasting
}

\author{
	Nikolaos Passalis, Anastasios Tefas, Juho Kanniainen, Moncef Gabbouj, and Alexandros Iosifidis
	\thanks{Nikolaos Passalis, Juho Kanniainen and Moncef Gabbouj are with the Faculty of Information Technology and Communication, Tampere University, Finland. Anastasios Tefas is with the School of Informatics, Aristotle University of Thessaloniki, Greece.  Alexandros Iosifidis is with the Department of Engineering, Electrical and Computer Engineering, Aarhus University, Denmark.  E-mail: nikolaos.passalis@tuni.fi, tefas@csd.auth.gr, juho.kanniainen@tuni.fi, moncef.gabbouj@tuni.fi, alexandros.iosifidis@eng.au.dk
	}

}

\maketitle

\begin{abstract}
	Deep Learning (DL) models can be used to tackle time series analysis tasks with great success. However, the performance of DL models can degenerate rapidly if the data are not appropriately normalized. This issue is even more apparent when DL is used for financial time series forecasting tasks, where the non-stationary and multimodal nature of the data  pose significant challenges and severely affect the performance of DL models. In this work, a simple, yet effective, neural layer, that is capable of adaptively normalizing the input time series, while taking into account the distribution of the data, is proposed. The proposed layer is trained in an end-to-end fashion using back-propagation and leads to significant performance improvements compared to other evaluated normalization schemes. The proposed method differs from traditional normalization methods since it learns how to perform normalization for a given task instead of using a fixed normalization scheme.   At the same time, it can be directly applied to any new time series without requiring re-training. The effectiveness of the proposed method is demonstrated using a large-scale limit order book dataset, as well as a load forecasting dataset.
\end{abstract}

	\section{Introduction}
Forecasting time series is an increasingly important topic, with several applications in various domains~\cite{kim2003financial,miranian2013developing, yan2012toward,ak2016two,deng2017deep, makinen2018forecasting,passalis2018temporal, nousi2019machine}. Many of these tasks are nowadays tackled using powerful deep learning (DL) models~\cite{greff2017lstm, kuremoto2014time, Fin_LSTM,gharehbaghi2018deep, tran2018temporal}, which often lead to state-of-the-art results outperforming the previously used methods. However, applying deep learning models to time series is challenging due to the non-stationary and multimodal nature of the data. This issue is even more apparent for financial time series, since financial data can exhibit significantly different behavior over the time due to a number of reasons, e.g., market volatility.

To allow for training deep learning models with time series data, the data must be first appropriately normalized. Perhaps the most widely used normalization scheme for time series when using DL is the z-score normalization,  i.e., subtracting the mean value of the data and dividing by their standard deviation. However, z-score normalization is unable to efficiently handle non-stationary time series, since the statistics used for the normalization are fixed both during the training and inference. Several recent works attempt to tackle this issue either by employing more sophisticated normalization schemes~\cite{ogasawara2010adaptive, nayak2014impact, shao2015self} or by using carefully handcrafted stationary features~\cite{tsantekidis2018using}.  Even though these approaches can indeed lead to slightly better performance when used to train deep learning models, they exhibit significant drawbacks, since they are largely based on heuristically-designed normalization/feature extraction 
schemes, e.g., using price change percentages instead of absolute prices, etc., while there is no actual guarantee that the designed scheme will be indeed be optimal for the task at hand.

To overcome these limitations, we propose a Deep Adaptive Input Normalization (DAIN) layer that is capable of a) learning how the data should be normalized and b) \textit{adaptively} changing the applied normalization scheme during inference, according to the distribution of the measurements of the current time series, allowing for effectively handling non-stationary and multimodal data. The proposed scheme is straightforward to implement, can be directly trained along with the rest of the parameters of a deep model in an \textit{end-to-end} fashion using back-propagation and can lead to impressive improvements in the forecasting accuracy.  Actually, as we experimentally demonstrate in Section~\ref{section:evaluation}, the proposed method allows for directly training deep learning models without applying any form of normalization to the data, since the raw time series is directly fed to the used deep learning model. 

\begin{figure*}
	\begin{center}
		\includegraphics[width=0.99\linewidth]{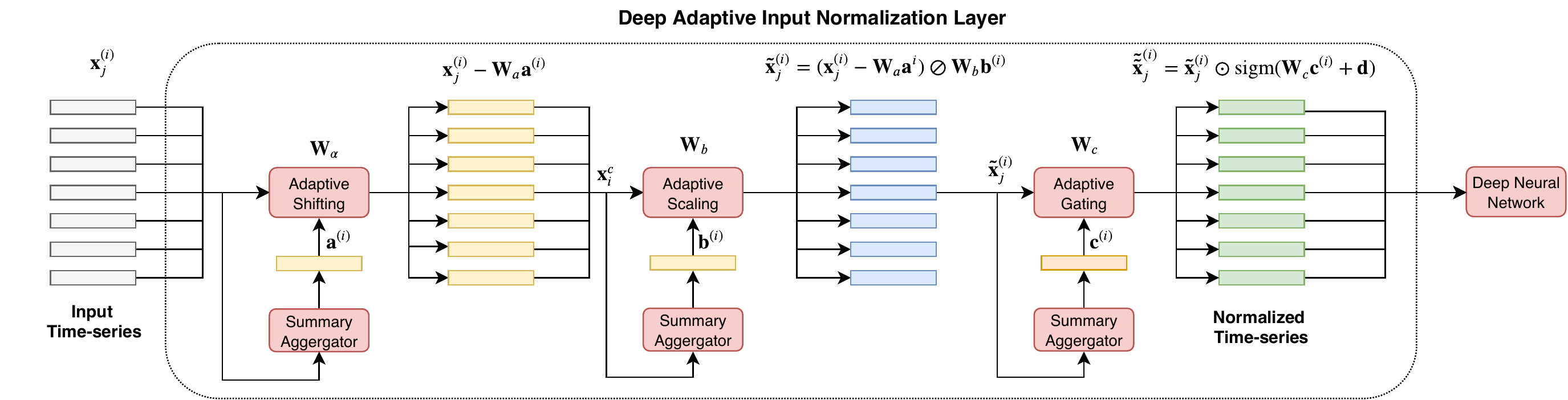}
	\end{center}
	\caption{Architecture of the proposed Deep Adaptive Input Normalization Layer (DAIN)}
	\label{fig:proposed}
\end{figure*}

The main contribution of this work is the proposal of a deep learning layer that learns how the data should be normalized according to their distribution instead of using fixed normalization schemes. To this end, the proposed layer is formulated as a series of three sublayers, as show in Fig.~\ref{fig:proposed}. The first layer is responsible for \textit{shifting} the data into the appropriate region of the feature space (centering), while the second layer is responsible for linearly \textit{scaling} the data in order to increase or reduce their variance (standardization) The third layer is responsible for performing \textit{gating}, i.e., non-linearly suppressing features that are irrelevant or not useful for the task at hand.  Note that the aforementioned process is adaptive, i.e., the applied normalization scheme depends on the actual time series that is fed to the network, and it is also trainable, i.e., the way the proposed layers behave is adapted to the task at hand using back-propagation.  The effectiveness of the proposed approach is evaluated using a large-scale limit order book  dataset that consists of 4.5~million limit orders~\cite{ntakaris2018mid}, as well as  a load forecasting dataset~\cite{hebrail2012individual}. An open-source implementation of the proposed method, along with code to reproduce the experiments conducted in this paper, are available at \url{https://github.com/passalis/dain}.

To the best of our knowledge this is the first time that an adaptive and trainable normalization scheme is proposed and effectively used in deep neural networks. In contrast to regular normalization approaches, e.g., z-score normalization, the proposed method a) \textit{learns} how to perform normalization for the task at hand (instead of using some fixed statistics calculated beforehand) and b) effectively exploits information regarding all the available features (instead of just using information for each feature of the time series separately). 
The proposed approach is also related to existing normalization approaches for deep neural networks, e.g., batch normalization~\cite{ioffe2015batch}, instance normalization~\cite{huang2017arbitrary}, layer normalization~\cite{ba2016layer} and group normalization~\cite{wu2018group}. However, these approaches are not actually designed for normalizing the input data and, most importantly, they are merely based on the statistics that are calculated during the training/inference, instead of \textit{learning} to dynamically normalize the data. It is worth noting that {it is not straightforward to use non-linear neural layers for adaptively normalizing the data}, since these layers usually require normalized data in the first place in order to function correctly. In this work, this issue is addressed by first using two robust and carefully initialized linear layers to estimate how the data should be centered and scaled, and then performing gating on the data using a non-linear layer that operates on the output of the previous two layers, effectively overcoming this limitation.

The rest of the paper is structured as follows. First, the proposed method is analytically described in Section~\ref{section:proposed}.  Then, an extensive experimental evaluation is provided in Section~\ref{section:evaluation}, while conclusions are drawn in Section~\ref{section:conclusions}.

\section{Deep Adaptive Input Normalization}
\label{section:proposed}
Let  $\{\mathbf{X}^{(i)} \in \mathbb{R}^{d \times L}; i = 1,...,N\}$ be a collection of $N$ time series, each of them composed of $L$ $d$-dimensional measurements (or features). 
The notation $\mathbf{x}^{(i)}_{j} \in \mathbb{R}^d$, $j = 1, 2, \dots, L$ is used to refer to the $d$ features observed at time point $j$ in time series $i$. Perhaps the most widely used form of normalization is to perform z-score scaling on each of the features of the time series. Note that if the data were not generated by a unimodal Gaussian distribution, then using the mean and standard deviation can lead to sub-optimal results, especially if the statistics around each mode significantly differ from each other. In this case, it can be argued that data should be normalized in an \textit{mode-aware} fashion, allowing for forming a common representation space that does not depend on the actual mode of the data. Even though this process can discard useful information, since the mode can provide valuable information for identifying each time series, at the same time it can hinder the generalization abilities of a model, especially for forecasting tasks. This can be better understood by the following example: assume two tightly connected companies with very different stock prices, e.g., 1\$ and 100\$ respectively. Even though the price movements can be very similar for these two stocks, the trained forecasting models will only observe very small variations around two very distant modes (if the raw time series are fed to the model). As a result, discarding the mode information completely can potentially improve the ability of the model to handle such cases, as we will further demonstrate in Section~\ref{section:evaluation}, since the two stocks will have very similar representations.


The goal of the proposed method is to \textit{learn} how the measurements $\mathbf{x}^{(i)}_{j}$ should be normalized by appropriately shifting and scaling them:
\begin{equation}\label{eq:1}
\mathbf{\tilde{x}}^{(i)}_j = \left(\mathbf{x}^{(i)}_j - \bm{\alpha}^{(i)} \right) \oslash \bm{\beta}^{(i)},
\end{equation}
where $\oslash$ is the Hadamard (entrywise) division operator. Note that global z-score normalization is a special case with $\bm{\alpha}^{(i)} = \bm{\alpha} = \left[\mu_1, \mu_2, \dots, \mu_d \right]$ and $\bm{\beta}^{(i)} = \bm{\beta} = \left[\sigma_1, \sigma_2, \dots, \sigma_d \right]$, where $\mu_k$ and $\sigma_k$ refer to the global average and standard deviation of the $k$-th input feature:
\[
\mu_k = \frac{1}{NL}\sum_{i=1}^N \sum_{j=1}^L x^{(i)}_{j,k}, \text{\quad}
\sigma_k = \sqrt{\frac{1}{NL}\sum_{i=1}^N \sum_{j=1}^L \left(x^{(i)}_{j,k} - \mu_k\right)^2}.
\]
However,  as it was already discussed, the obtained estimations for $\bm{\alpha}$ and $\bm{\beta}$ might not be the optimal for normalizing every possible measurement vector, since the distribution of the data might significantly drift, invalidating the previous choice for these parameters. This issue becomes even more apparent when the data are multimodal, e.g., when training model using time series  data from different stocks that exhibit significantly different behavior (price levels, trading frequency, etc.). To overcome these limitations we propose to dynamically estimate these quantities and \textit{separately} normalize each time series by \textit{implicitly} estimating the distribution from which each measurement was generated. Therefore, in this work, we propose normalizing each time series 
so that $\bm{\alpha}$ and $\bm{\beta}$ are {\em learned and depend on the current input}, instead of being the global averages calculated using the whole dataset.

The proposed architecture is summarized in Fig.~\ref{fig:proposed}. First a \textit{summary representation} of the time series is extracted by averaging all the $L$ measurements:
\begin{equation}
\mathbf{a}^{(i)} = \frac{1}{L} \sum_{j=1}^{L} \mathbf{x}^{(i)}_j \in \mathbb{R}^d.
\end{equation}
This representation provides an initial estimation for the mean of the current time series and, as a result, it can be used to estimate the distribution from which the current time series was generated, in order to appropriately modify the normalization procedure. Then, the shifting operator $\bm{\alpha}^{(i)}$ is defined using a linear transformation of the extracted summary representation as:
\begin{equation}
\label{eq:2}
\bm{\alpha}^{(i)}=\mathbf{W}_a \mathbf{a}^{(i)} \in \mathbb{R}^{d},
\end{equation}
where $\mathbf{W}_a  \in  \mathbb{R}^{d \times d}$ is the weight matrix of the first neural layer, which is responsible for shifting the measurements across each dimension. Employing a linear transformation layer ensures that the proposed method will be able to handle data that are not appropriately normalized (or even not normalized at all), allowing for training the proposed model in an end-to-end fashion without having to deal with stability issues, such as saturating the activation functions.  This layer is called \textit{adaptive shifting layer}, since it estimates how the data must be shifted before feeding them to the network. Note that this approach allows for exploiting possible correlations between different features to perform more robust normalization. 

After centering the data using the process described in~(\ref{eq:2}), the data must be appropriately scaled using the scaling operator $\bm{\beta}^{(i)}$. To this end, we calculate an updated summary representation that corresponds to the standard deviation of the data as:
\begin{equation}
b\textbf{}^{(i)}_k = \sqrt{\frac{1}{L} \sum_{j=1}^L \left(x^{(i)}_{j,k} - \alpha^{(i)}_{k} \right)^2}, \text{\quad} k = 1, 2, \dots, d.
\end{equation}
Then, the scaling function can be similarly defined as a linear transformation of this summary representation allowing for scaling each of the shifted measurements:
\begin{equation}
\label{eq:3}
\bm{\beta}^{(i)} = \mathbf{W}_b \mathbf{b}^{(i)} \in \mathbb{R}^{d},
\end{equation}
where $\mathbf{W}_b \in  \mathbb{R}^{d \times d}$ is the weight matrix the scaling layer. This layer is called \textit{adaptive scaling layer}, since it estimates how the data must be scaled before feeding them to the network. Also, note that this process corresponds to scaling the data according to their variance, as performed with z-score normalization.

Finally, the data are fed to an \textit{adaptive gating layer}, which is capable of suppressing features that are not relevant or useful for the task as hand as:
\begin{equation}
\label{eq:5}
\mathbf{\tilde{\tilde{x}}}^{(i)}_j = \mathbf{\tilde{x}}^{(i)}_j \odot \bm{\gamma}^{(i)},
\end{equation}
where $\odot$ is Hadamard (entrywise) multiplication operator and
\begin{equation}
\label{eq:6}
\bm{\gamma}^{(i)} = \text{sigm}(\mathbf{W}_c \mathbf{c}^{(i)} +\mathbf{d}) \in \mathbb{R}^{d},
\end{equation}
$\text{sigm}(x) = 1 / (1 + \exp (-x) )$ is the sigmoid function,  $\mathbf{W}_c \in  \mathbb{R}^{d \times d}$ and $\mathbf{d} \in \mathbb{R}^{d}$ are the parameters of the gating layer, and   $\mathbf{c}^{(i)}$ is a third summary representation calculated as:
\begin{equation}
\mathbf{c}^{(i)} = \frac{1}{L} \sum_{j=1}^{L} \mathbf{\tilde{x}}^{(i)}_j \in \mathbb{R}^d.
\end{equation}
Note that in contrast with the previous layers, this layer is non-linear and it is capable of suppressing the normalized features. In this way, features that are not relevant to the task at hand or can harm the generalization abilities of the network, e.g., features with excessive variance, can be appropriate filtered before being fed to the network. Overall, $\bm{\alpha}^{(i)}, \bm{\beta}^{(i)}, \bm{\gamma}^{(i)}$ are dependent on current 'local' data on window $i$ and the 'global' estimates of $\mathbf{W}_{a}, \mathbf{W}_{b}, \mathbf{W}_{c}, \mathbf{d}$ that are trained using multiple samples on time-series, $\{\mathbf{X}^{(i)} \in \mathbb{R}^{d \times L}; i = 1,...,M\}$, where $M$ is the number of samples in the training data.

The output of the proposed normalization layer, which is called Deep Adaptive Input Normalization (DAIN), can be obtained simply by feed-forwarding through its three layers, as shown in Fig.~\ref{fig:proposed}, while the parameters of the layers are kept fixed during the inference process. Therefore, no additional training is required during inference. All the parameters of the resulting deep model can be directly learned in an end-to-end fashion using gradient descent:
\begin{dmath}
	\small
	\Delta \Big( \mathbf{W}_{a}, \mathbf{W}_{b}, \mathbf{W}_{c}, \mathbf{d}, \mathbf{W} \Big) = - \eta\Big( \eta_a \frac{\partial \mathcal{L}}{\partial \mathbf{W}_{a}}, \eta_b \frac{\partial \mathcal{L}}{\partial \mathbf{W}_{b}},\eta_c \frac{\partial \mathcal{L}}{\partial \mathbf{W}_{c}},\eta_c \frac{\partial \mathcal{L}}{\partial \mathbf{d}}, \frac{\partial \mathcal{L}}{\partial \mathbf{W}} \Big)
\end{dmath}
where $\mathcal{L}$ denotes the loss function used for training the network and $\mathbf{W}$ denotes the weights of the neural network that follows the proposed layer.  Therefore, the proposed normalization scheme can be used on top of every deep learning network and the resulting architecture can be trained using the regular back-propagation algorithm, as also experimentally demonstrated in Section~\ref{section:evaluation}. Note that separate learning rates are used for the parameters of each sub-layer, i.e., $\eta_a$,  $\eta_b$ and $\eta_c$. This was proven essential to ensure the smooth convergence of the proposed method due to the enormous differences in the resulting gradients between the parameters of the various sub-layers.

\begin{table}[ht!]
	\caption{Ablation study using the FI-2010 dataset}
	\label{table:ablation-study}
	
	\begin{center}
		\begin{tabular}{ll||ccccc}
			\textbf{ Method}  & \textbf{Model} &   \textbf{Macro F1} & \textbf{Cohen's} $\bm{\kappa}$ \\
			\midrule
			No norm.  & MLP & $ 12.71 \pm 13.22 $ & $ 0.0010 \pm 0.0014 $\\
			z-score norm. & MLP & $ 53.76 \pm 0.99 $ & $ 0.3059 \pm 0.0157 $ \\
			Sample avg norm. & MLP  & $ 41.80 \pm 3.58 $ & $ 0.1915 \pm 0.0284 $ \\
			\midrule
			Batch Norm. & MLP &  $ 52.72 \pm 1.94 $ & $ 0.2893 \pm 0.0264 $ \\
			Instance Norm.  & MLP &  $ 59.13 \pm 2.94 $ & $ 0.3717 \pm 0.0406 $ \\
			\midrule
			DAIN (1) & MLP & $ 57.37 \pm 3.16 $ & $ 0.3536 \pm 0.0417 $ \\
			DAIN (1+2) & MLP  & $ 66.71 \pm 2.02 $ & $ 0.4896 \pm 0.0289 $ \\
			DAIN (1+2+3) & MLP&  $\mathbf{66.92 \pm 1.70}$ & $ \mathbf{0.4934 \pm 0.0238}$\\			
			\midrule
			\midrule
			
			No norm.  & CNN &  $ 12.61 \pm 12.89 $ & $ 0.0003 \pm 0.0006 $\\
			z-score norm. & CNN&  $ 50.94 \pm 1.12 $ & $ 0.2570 \pm 0.0184 $ \\
			Sample avg norm. & CNN &  $ 53.49 \pm 3.38 $ & $ 0.2934 \pm 0.0458 $ \\
			\midrule
			Batch Norm. & CNN &  $ 45.89 \pm 3.40 $ & $ 0.1833 \pm 0.0517 $ \\
			Instance Norm.  & CNN & $ 57.05 \pm 1.61 $ & $ 0.3396 \pm 0.0219 $ \\
			\midrule
			DAIN (1) & CNN & $ 59.79 \pm 1.46 $ & $ 0.3838 \pm 0.0199 $ \\
			DAIN (1+2) & CNN &  $ 61.91 \pm 3.65 $ & $ 0.4136 \pm 0.0574 $ \\
			DAIN (1+2+3) & CNN &$\mathbf{63.02 \pm 2.40}$ & $ \mathbf{ 0.4327 \pm 0.0358}$\\			
			
			\midrule
			\midrule
			
			No norm.  & RNN & $ 31.61 \pm 0.40 $ & $ 0.0075 \pm 0.0024 $ \\
			z-score norm. & RNN &  $ 52.29 \pm 2.10 $ & $ 0.2789 \pm 0.0295 $ \\
			Sample avg norm. & RNN & $ 49.47 \pm 2.73 $ & $ 0.2277 \pm 0.0403 $ \\
			\midrule
			Batch Norm. & RNN &  $ 51.42 \pm 1.05 $ & $ 0.2668 \pm 0.0147 $ \\
			Instance Norm.  & RNN &  $ 54.01 \pm 3.41 $ & $ 0.2979 \pm 0.0448 $\\
			\midrule
			DAIN (1) & RNN &  $ 55.34 \pm 2.88 $ & $ 0.3164 \pm 0.0412 $\\
			DAIN (1+2) & RNN &  $\mathbf{64.21 \pm 1.47} $ & $\mathbf{0.4501 \pm 0.0197}$  \\
			DAIN (1+2+3) & RNN  & $ 63.95 \pm 1.31 $ & $ 0.4461 \pm 0.0168 $\\	
		\end{tabular}
	\end{center}
\end{table}

\begin{table*}[ht!]
	\caption{Evaluation results using the FI-2010 dataset }
	\label{table:res-main}
	
	\begin{center}
		\begin{tabular}{lll|cccc}
			\textbf{Normalization Method} & \textbf{Model} & \textbf{Horizon}  &  \textbf{Macro Precision} &  \textbf{Macro Recall } &  \textbf{Macro F1 score} & \textbf{Cohen's} $\bm{\kappa}$ \\
			\midrule
			z-score & MLP & 10 & $ 50.50 \pm 2.03$ & $ 65.31 \pm 4.29$ & $ 54.65 \pm 2.34 $ & $ 0.3206 \pm 0.0351 $\\
			Instance Normalization & MLP & 10 & $ 54.89 \pm 2.88$ & $ 70.08 \pm 2.90$ & $ 59.67 \pm 2.26 $ & $ 0.3827 \pm 0.0316 $\\
			DAIN & MLP & 10 & $\mathbf{65.67 \pm 2.26}$ & $\mathbf{71.58 \pm 1.21}$ & $\mathbf{68.26 \pm 1.67}$ & $\mathbf{0.5145 \pm 0.0256}$\\
			\midrule
			z-score & MLP & 20 & $ 52.08 \pm 2.33$ & $ 64.41 \pm 3.58$ & $ 54.66 \pm 2.68 $ & $ 0.3218 \pm 0.0361 $ \\
			Instance Normalization & MLP & 20 & $ 57.34 \pm 2.67$ & $\mathbf{70.77 \pm 2.32}$ & $ 61.12 \pm 2.33 $ & $ 0.3985 \pm 0.0305 $\\
			DAIN & MLP & 20 & $\mathbf{62.10 \pm 2.09}$ & $ 70.48 \pm 1.93$ & $\mathbf{65.31 \pm 1.62}$ & $\mathbf{0.4616 \pm 0.0237}$\\
			\midrule
			\midrule
			
			z-score & RNN & 10  & $ 53.73 \pm 2.42$ & $ 54.63 \pm 2.88$ & $ 53.85 \pm 2.66 $ & $ 0.3018 \pm 0.0412 $ \\
			Instance Normalization & RNN & 10 &  $ 58.68 \pm 2.51$ & $ 57.72 \pm 3.90$ & $ 57.85 \pm 2.23 $ & $ 0.3546 \pm 0.0346 $ \\
			DAIN & RNN & 10 & $\mathbf{61.80 \pm 3.19}$ & $\mathbf{70.92 \pm 2.53}$ & $\mathbf{65.13 \pm 2.37}$ & $\mathbf{0.4660 \pm 0.0363}$\\
			\midrule
			z-score & RNN & 20  & $ 53.05 \pm 2.28$ & $ 55.79 \pm 2.43$ & $ 53.97 \pm 2.31 $ & $ 0.2967 \pm 0.0353 $ \\
			Instance Normalization & RNN & 20 & $ 58.13 \pm 2.39$ & $ 60.11 \pm 2.24$ & $ 58.75 \pm 1.53 $ & ${0.3588 \pm 0.0234}$\\
			DAIN & RNN & 20 & $\mathbf{59.16 \pm 2.21}$ & $\mathbf{68.51 \pm 1.54}$ & $ \mathbf{62.03 \pm 2.20}$ & $ \mathbf{0.4121 \pm 0.0331}$\\
		\end{tabular}
	\end{center}
\end{table*}

\begin{table}[ht!]
	\caption{Evaluation results using the Household Power Consumption dataset}
	\label{table:power}
	\begin{center}
		\begin{tabular}{ll|c}
			\textbf{Normalization Method} & \textbf{Model} & \textbf{Accuracy (\%)}  \\
			\midrule
			None & MLP &  71.57 \\			
			z-score & MLP &  75.39 	\\
			Instance Normalization & MLP & 77.93 \\
			DAIN & MLP & $\mathbf{78.83}$\\	
			\midrule
			None & RNN &  77.16\\			
			z-score & RNN &  77.22 	\\
			Instance Normalization & RNN & 77.25 \\
			DAIN & RNN & $\mathbf{78.59}$\\

		\end{tabular}
	\end{center}
\end{table}


\section{Experimental Evaluation}
\label{section:evaluation}

For evaluating the proposed method a challenging large-scale dataset (FI-2010), that contains limit order book data, was employed~\cite{ntakaris2018mid}. The data were collected from 5 Finnish companies traded in the Helsinki Exchange (operated by Nasdaq Nordic) and the ten highest and ten lowest ask/bid order prices were measured. The data were gathered over a period of 10 business days from 1st June 2010 to 14th June 2010. Then, the pre-processing and feature extraction pipeline proposed in~\cite{kercheval2015modelling} was employed for processing the 4.5~million limit orders that were collected, leading to a total of 453,975 144-dimensional feature vectors that were extracted. 

We also followed the anchored evaluation setup that was proposed in~\cite{tomasini2011trading}. According to this setup the time series that were extracted from the first day were used to train the model and the data from the second day were used for evaluating the method. Then, the first two days were employed for training the methods, while the data from the next day were used for the evaluation. This process was repeated 9 times, i.e., one time for each of the days available in the dataset (except from the last one, for which no test data are available). The performance of the evaluated methods was measured using the macro-precision, macro-recall, macro-F1 and Cohen's $\kappa$. Let $TP_c$, $FP_c$, $TN_c$ and $FN_c$ be the true positives, false positives, true negatives and false negatives of class $c$. The precision of a class is defined as $prec_c = TP_c / (TP_c + FP_c)$, the recall as $recall_c = TP_c / (TP_c + FN_c)$, while the F1 score for a class $c$ is calculated as the harmonic mean of the precision and the recall: $F1_c = 2\cdot(prec_c \cdot recall_c)/ (prec_c + recall_c)$. These metrics are calculated for each class separately and then averaged (macro-averaging). Finally, using the Cohen's $\kappa$ metric allows for evaluating the agreement between two different sets of annotations, while accounting for the possible random agreements. The mean and standard deviation values over the anchored splits are reported. The trained models were used for predicting the direction of the average mid price (up, stationary or down) after 10 and 20 time steps, while a stock was considered stationary if the change in the mid price was less than to $0.01\%$ (or $0.02\%$ for the prediction horizon of 20 time steps).

Three different neural network architectures were used for the evaluation: a Multilayer Perceptron (MLP)~\cite{nousi2018machine}, a Convolutional Neural Network (CNN)~\cite{cui2016multi,tsantekidis2017forecasting} and a Recurrent Neural Network (RNN) composed of Gated Recurrent Units~\cite{chung2014empirical}. All the evaluated models receive as input the 15 most recent measurement (feature) vectors extracted from the time series and predict the future price direction. For the MLP the measurements are flattened into a constant length vector with $15\times 144= 2,160$ measurements, maintaining in this way the temporal information of the time series. The MLP is composed of one fully connected hidden layer with 512 neurons (the ReLU activation function is used~\cite{glorot2011deep}) followed by a fully connected layer with 3 output neurons (each one corresponding to one of the predicted categories). Dropout with rate of 0.5\% is used after the hidden layer~\cite{srivastava2014dropout}. The CNN is composed of a 1-D convolution layer with 256 filters and kernel size of 3, followed by two fully connected layers with the same architectures as in the employed MLP. The RNN is composed of a GRU layer with 256 hidden units, followed by two fully connected layers with the same architectures as in the employed MLP. The networks were trained using the cross-entropy loss.

First, an ablation study was performed to identify the effect of each normalization sub-layer on the performance of the proposed method. The results are reported in Table~\ref{table:ablation-study}. The notation ``DAIN (1)'' is used to refer to applying only (\ref{eq:2}) for the normalization process, the notation ``DAIN (1+2)'' refers to using the first two layers for the normalization process, while the notation ``DAIN (1+2+3)'' refers to using all the three normalization layers. The optimization ran for 20 epochs over the training data, while for the evaluation the first 3 days (1, 2 and 3) were employed using the anchored evaluation scheme that was previously described. The proposed method is also compared to a) not applying any form of normalization to the data (``No norm.''), b) using z-score normalization, c) subtracting the average measurement vector from each time series (called ``Sample avg norm.'' in Table~\ref{table:ablation-study}), d) using the Batch Normalization~\cite{ioffe2015batch} and e) Instance Normalization layers~\cite{huang2017arbitrary} directly on the input data. Note that Batch Normalization and Instance Normalization were not originally designed for normalizing the input data. However, they can be used for this task, providing an additional baseline.  All the three models (MLP, CNN and RNN) were used for the evaluation, while the models were trained for 20 training epochs over the data. Furthermore, the data were sampled with probability inversely proportional to their class frequency, to ensure that each class is equally represented during the training. Thus, data from the less frequent classes were sampled more frequently and vice versa. For all the conducted experiments of the ablation study the prediction horizon was set for the next 10 time steps.

Several conclusions can be drawn from the results reported in Table~\ref{table:ablation-study}. First, using some form of normalization is essential for ensuring that the models will be successfully trained, since using no normalization leads to $\kappa$ values around 0 (random agreement). Using either z-score normalization or performing sample-based normalization seems to work equally well.  Batch Normalization yields performance similar to the z-score normalization, as expected, while Instance Normalization improves the performance over all the other baseline normalization approaches.  When the first layer of the proposed DAIN method is applied (adaptive shifting) the performance of the model over the fixed normalization approaches increases (relative improvement) by more than 15\%  for the MLP,  30\%  for the CNN and 13\% for the RNN (Cohen's $\kappa$), highlighting that \textit{learning} how to adaptively shift each measurement vector, based on the distribution from which the sample was generated, can indeed lead to significant improvements. Note that the adaptive shifting layer directly receives the raw data, without any form of normalization, and yet it manages to learn how they should be normalized in order to successfully train the rest of the network. A key ingredient for this was to a) avoid using any non-linearity in the shifting process (that could possibly lead to saturating the input neurons) and b) appropriately initializing the shifting layer, as previously described.  Using the additional adaptive scaling layer, that also scales each measurement separately, further improves the performance for all the evaluated model. Finally, the adaptive gating layer improves the performance for the MLP and CNN (average relative improvement of about 2.5\%). However, it does not further improve the performance of the GRU. This can be explained since GRUs already incorporate various gating mechanisms that can provide the same functionality as the employed third layer of DAIN.

Then, the models were evaluated using the full training data (except from the first day which was used to tune the hyper-parameters of the proposed method) and two different prediction horizons (10 and 20 time steps). The experimental results are reported in Table~\ref{table:res-main} using the two best performing models  (MLP and RNN). Again, no other form of normalization, e.g., z-score, etc., was employed for the model that uses the proposed (full) DAIN layer and the Instance Normalization layer. Using Instance Normalization leads to better performance over the plain z-score normalization. However, employing the proposed  method again significantly improves the obtained results over the rest of the evaluated methods for both models.

Finally, the proposed method was also evaluated on an additional dataset, the Household Power Consumption dataset~\cite{hebrail2012individual}. The forecasting task used for this dataset was to predict whether the average power consumption of a household will increase or decrease the next 10 minutes, compared to the previous 20 minutes (a 90\%-10\% training/testing split was employed for the evaluation). The same MLP and RNN architectures as before were used for the conducted experiments, while 20 7-dimensional feature vectors with various measurements (one feature vector for each minute), were fed to the models. The results of the experimental evaluation are reported in Table~\ref{table:power}. Again, the proposed method leads to significant improvements over the three other evaluated methods. Also, note that even through the GRU model leads to significantly better results when simpler normalization methods are used, e.g., z-score, it achieves almost the same performance with the MLP when the proposed DAIN layer is used.

We also performed one additional experiment to evaluate the ability of the proposed approach to withstand distribution shifts and/or handle heavy-tailed datasets. More specifically, all the measurements fed to the model during the evaluation were shifted (increased) by adding 3 times their average value (except of the voltage measurements). This led to a decrease of classification performance from 75.39\% to 56.56\% for the MLP model trained with plain z-score normalization. On the other hand, the proposed method was only slightly affected: the classification accuracy was reduced less than 0.5\% (from 78.59\% to 78.21\%).

\textbf{Hyper-parameters:} The learning hyper-parameters were tuned for the FI-2010 dataset using a simple line search procedure (the first day of the dataset was used for the evaluation).  The base learning rate was set to $\eta=10^{-4}$, while the learning rates for the sub-layers were set as follows: $\eta_a=10^{-6}/10^{-2}/10^{-2}$, $\eta_b=10^{-3}/10^{-9}/10^{-8}$,  and  $\eta_c = 10/10/10$ (MLP/CNN/RNN respectively). For the household power consumption dataset  the learning rates were set to $\eta_a=10^{-5}$, $\eta_b=10^{-2}$,  and  $\eta_c = 10$. The weights of the adaptive shifting and adaptive scaling layers were initialized to the identity matrix, i.e., $\mathbf{W}_{a} = \mathbf{W}_{b} = \mathbf{I}_{d\times d}$, while the rest of the parameters were randomly initialized by drawing the weights from a normal distribution. The RMSProp algorithm was used for optimizing the resulting deep architecture~\cite{tieleman2012lecture}.

\section{Conclusions}
\label{section:conclusions}
A deep adaptive normalization method, that can be trained in an end-to-end fashion, was proposed in this paper. The proposed method is easy to implement, while at the same allows for directly using the raw time series data. The ability of the proposed method to improve the forecasting performance was evaluated using three different deep learning models and two time series forecasting datasets. The proposed method consistently outperformed all the other evaluated normalization approaches. 

There are several interesting future research direction. First, alternative and potentially stabler learning approaches, e.g., multiplicative weight updates, can be employed for updating the parameters of the DAIN layer reducing the need of carefully fine-tuning the learning rate for each sub-layer separately. Furthermore, more advanced aggregation methods can also be used for extracting the summary representation, such as extending the Bag-of-Features model~\cite{passalis2018training} to extract representations from time-series~\cite{passalis2019temporal}. Also, in addition to z-score normalization, min-max normalization, mean normalization, and scaling to unit length can be also expressed as special cases in the proposed normalization scheme, providing, among others, different initialization points. Finally, methods that can further enrich the extracted representation with mode information (which is currently discarded) can potentially further improve the performance of the models.

\bibliographystyle{plain}
\bibliography{mybib}

\begin{thebibliography}{10}

\bibitem{ak2016two}
Ronay Ak, Olga Fink, and Enrico Zio.
\newblock Two machine learning approaches for short-term wind speed time-series
  prediction.
\newblock {\em IEEE Trans. on Neural Networks and Learning Systems},
  27(8):1734--1747, 2016.

\bibitem{ba2016layer}
Jimmy~Lei Ba, Jamie~Ryan Kiros, and Geoffrey~E Hinton.
\newblock Layer normalization.
\newblock {\em arXiv preprint arXiv:1607.06450}, 2016.

\bibitem{chung2014empirical}
Junyoung Chung, Caglar Gulcehre, KyungHyun Cho, and Yoshua Bengio.
\newblock Empirical evaluation of gated recurrent neural networks on sequence
  modeling.
\newblock {\em arXiv preprint arXiv:1412.3555}, 2014.

\bibitem{cui2016multi}
Zhicheng Cui, Wenlin Chen, and Yixin Chen.
\newblock Multi-scale convolutional neural networks for time series
  classification.
\newblock {\em arXiv preprint arXiv:1603.06995}, 2016.

\bibitem{deng2017deep}
Yue Deng, Feng Bao, Youyong Kong, Zhiquan Ren, and Qionghai Dai.
\newblock Deep direct reinforcement learning for financial signal
  representation and trading.
\newblock {\em IEEE Trans. on Neural Networks and Learning Systems},
  28(3):653--664, 2017.

\bibitem{gharehbaghi2018deep}
Arash Gharehbaghi and Maria Lind{\'e}n.
\newblock A deep machine learning method for classifying cyclic time series of
  biological signals using time-growing neural network.
\newblock {\em IEEE Trans. on Neural Networks and Learning Systems},
  29(9):4102--4115, 2018.

\bibitem{glorot2011deep}
Xavier Glorot, Antoine Bordes, and Yoshua Bengio.
\newblock Deep sparse rectifier neural networks.
\newblock In {\em Proc. of the Int. Conf. on Artificial Intelligence and
  Statistics}, pages 315--323, 2011.

\bibitem{greff2017lstm}
Klaus Greff, Rupesh~K Srivastava, Jan Koutn{\'\i}k, Bas~R Steunebrink, and
  J{\"u}rgen Schmidhuber.
\newblock Lstm: A search space odyssey.
\newblock {\em IEEE Trans. on Neural Networks and Learning Systems},
  28(10):2222--2232, 2017.

\bibitem{hebrail2012individual}
G~H{\'e}brail and A~B{\'e}rard.
\newblock Individual household electric power consumption data set.
\newblock {\em {\'E}. d. France, Ed., ed: UCI Machine Learning Repository},
  2012.

\bibitem{huang2017arbitrary}
Xun Huang and Serge~J Belongie.
\newblock Arbitrary style transfer in real-time with adaptive instance
  normalization.
\newblock In {\em Proc. of the Int. Conf. on Computer Vision}, pages
  1510--1519, 2017.

\bibitem{ioffe2015batch}
Sergey Ioffe and Christian Szegedy.
\newblock Batch normalization: Accelerating deep network training by reducing
  internal covariate shift.
\newblock In {\em Proc. of the Int. Conf. on Machine Learning}, pages 448--456,
  2015.

\bibitem{kercheval2015modelling}
Alec~N. Kercheval and Yuan Zhang.
\newblock Modelling high-frequency limit order book dynamics with support
  vector machines.
\newblock {\em Quantitative Finance}, 15(8):1315--1329, 2015.

\bibitem{kim2003financial}
Kyoung-jae Kim.
\newblock Financial time series forecasting using support vector machines.
\newblock {\em Neurocomputing}, 55(1-2):307--319, 2003.

\bibitem{kuremoto2014time}
Takashi Kuremoto, Shinsuke Kimura, Kunikazu Kobayashi, and Masanao Obayashi.
\newblock Time series forecasting using a deep belief network with restricted
  boltzmann machines.
\newblock {\em Neurocomputing}, 137:47--56, 2014.

\bibitem{makinen2018forecasting}
Milla M{\"a}kinen, Juho Kanniainen, Moncef Gabbouj, and Alexandros Iosifidis.
\newblock Forecasting of jump arrivals in stock prices: New attention-based
  network architecture using limit order book data.
\newblock {\em arXiv preprint arXiv:1810.10845}, 2018.

\bibitem{miranian2013developing}
Arash Miranian and Majid Abdollahzade.
\newblock Developing a local least-squares support vector machines-based
  neuro-fuzzy model for nonlinear and chaotic time series prediction.
\newblock {\em IEEE Trans. on Neural Networks and Learning Systems},
  24(2):207--218, 2013.

\bibitem{nayak2014impact}
SC~Nayak, BB~Misra, and HS~Behera.
\newblock Impact of data normalization on stock index forecasting.
\newblock {\em Int. Journal of. Computer Information Systems and Industrial
  Management Applications}, 6:357--369, 2014.

\bibitem{nousi2018machine}
Paraskevi Nousi, Avraam Tsantekidis, Nikolaos Passalis, Adamantios Ntakaris,
  Juho Kanniainen, Anastasios Tefas, Moncef Gabbouj, and Alexandros Iosifidis.
\newblock Machine learning for forecasting mid price movement using limit order
  book data.
\newblock {\em arXiv preprint arXiv:1809.07861}, 2018.

\bibitem{nousi2019machine}
Paraskevi Nousi, Avraam Tsantekidis, Nikolaos Passalis, Adamantios Ntakaris,
  Juho Kanniainen, Anastasios Tefas, Moncef Gabbouj, and Alexandros Iosifidis.
\newblock Machine learning for forecasting mid-price movements using limit
  order book data.
\newblock {\em IEEE Access}, 7:64722--64736, 2019.

\bibitem{ntakaris2018mid}
Adamantios Ntakaris, Martin Magris, Juho Kanniainen, Moncef Gabbouj, and
  Alexandros Iosifidis.
\newblock Benchmark dataset for mid-price prediction of limit order book data.
\newblock {\em Journal of Forecasting}, 37(8):852--866, 2018.

\bibitem{ogasawara2010adaptive}
Eduardo Ogasawara, Leonardo~C Martinez, Daniel De~Oliveira, Geraldo
  Zimbr{\~a}o, Gisele~L Pappa, and Marta Mattoso.
\newblock Adaptive normalization: A novel data normalization approach for
  non-stationary time series.
\newblock In {\em Proc. of the Int. Joint Conf. on Neural Networks}, pages
  1--8, 2010.

\bibitem{passalis2018training}
Nikolaos Passalis and Anastasios Tefas.
\newblock Training lightweight deep convolutional neural networks using
  bag-of-features pooling.
\newblock {\em IEEE Trans. on Neural Networks and Learning Systems}, 2018.

\bibitem{passalis2018temporal}
Nikolaos Passalis, Anastasios Tefas, Juho Kanniainen, Moncef Gabbouj, and
  Alexandros Iosifidis.
\newblock Temporal bag-of-features learning for predicting mid price movements
  using high frequency limit order book data.
\newblock {\em IEEE Trans. on Emerging Topics in Computational Intelligence},
  2018.

\bibitem{passalis2019temporal}
Nikolaos Passalis, Anastasios Tefas, Juho Kanniainen, Moncef Gabbouj, and
  Alexandros Iosifidis.
\newblock Temporal logistic neural bag-of-features for financial time series
  forecasting leveraging limit order book data.
\newblock {\em arXiv preprint arXiv:1901.08280}, 2019.

\bibitem{shao2015self}
Xiaofeng Shao.
\newblock Self-normalization for time series: a review of recent developments.
\newblock {\em Journal of the American Statistical Association},
  110(512):1797--1817, 2015.

\bibitem{srivastava2014dropout}
Nitish Srivastava, Geoffrey Hinton, Alex Krizhevsky, Ilya Sutskever, and Ruslan
  Salakhutdinov.
\newblock Dropout: a simple way to prevent neural networks from overfitting.
\newblock {\em The Journal of Machine Learning Research}, 15(1):1929--1958,
  2014.

\bibitem{tieleman2012lecture}
Tijmen Tieleman and Geoffrey Hinton.
\newblock {RMSProp}: Divide the gradient by a running average of its recent
  magnitude.
\newblock {\em COURSERA: Neural networks for machine learning}, 4(2):26--31,
  2012.

\bibitem{tomasini2011trading}
Emilio Tomasini and Urban Jaekle.
\newblock {\em Trading {S}ystems}.
\newblock Harriman House Limited, Hampshire, UK, 2011.

\bibitem{tran2018temporal}
Dat~Thanh Tran, Alexandros Iosifidis, Juho Kanniainen, and Moncef Gabbouj.
\newblock Temporal attention-augmented bilinear network for financial
  time-series data analysis.
\newblock {\em IEEE Trans. on Neural Networks and Learning Systems}, 2018.

\bibitem{tsantekidis2017forecasting}
Avraam Tsantekidis, Nikolaos Passalis, Anastasios Tefas, Juho Kanniainen,
  Moncef Gabbouj, and Alexandros Iosifidis.
\newblock Forecasting stock prices from the limit order book using
  convolutional neural networks.
\newblock In {\em Proc. of the IEEE Conf. on Business Informatics (CBI)}, pages
  7--12, 2017.

\bibitem{Fin_LSTM}
Avraam Tsantekidis, Nikolaos Passalis, Anastasios Tefas, Juho Kanniainen,
  Moncef Gabbouj, and Alexandros Iosifidis.
\newblock Using deep learning to detect price change indications in financial
  markets.
\newblock In {\em Proc. of the European Signal Processing Conf.}, pages
  2511--2515, 2017.

\bibitem{tsantekidis2018using}
Avraam Tsantekidis, Nikolaos Passalis, Anastasios Tefas, Juho Kanniainen,
  Moncef Gabbouj, and Alexandros Iosifidis.
\newblock Using deep learning for price prediction by exploiting stationary
  limit order book features.
\newblock {\em arXiv preprint arXiv:1810.09965}, 2018.

\bibitem{wu2018group}
Yuxin Wu and Kaiming He.
\newblock Group normalization, 2018.

\bibitem{yan2012toward}
Weizhong Yan.
\newblock Toward automatic time-series forecasting using neural networks.
\newblock {\em IEEE Trans. on Neural Networks and Learning Systems},
  23(7):1028--1039, 2012.

\end{thebibliography}

\end{document}